\documentclass[12pt,leqno]{article}
\usepackage{amsmath,amssymb,amsbsy,amsfonts,amsthm,latexsym,
         amsopn,amstext,amsxtra,euscript,amscd,amsthm}
\usepackage{graphicx}
\usepackage{authblk}

\date{\vspace{-5ex}}

\author[1]{R\'eal Tremblay\thanks{real.tremblay@phy.ulaval.ca}}
\author[2]{Nicolas Doyon\thanks{nicolas.doyon@mat.ulaval.ca}}
\author[1]{Julien Beaudoin-Bertrand\thanks{julien.b.bertrand@copl.ulaval.ca}}
\affil[1]{Centre d'optique, photonique et laser (COPL), D\'epartement de physique, g\'enie physique et optique, Universit\'e Laval, Qu\'ebec, G1V 0A6, Canada}
\affil[2]{D\'epartement de math\'ematiques et de statistiques, Universit\'e Laval, Qu\'ebec, G1V 0A6, Canada}
\title{TE-TM Electromagnetic modes and states in quantum physics}

\begin{document}
\maketitle
\begin{abstract}
	We propose, as another pedagogical approach, a quantification of the e.m. field better adapted in isolated systems, a quantification of e.m. fields in finite-spacetime which does not rely explicitly on the notion of photon, nor on the general application of gauge. Being based on the development of e.m. field in TE-TM modes and states, it obeys the limit conditions of a finite-spacetime which follows from the solution of an eigenvalue equation and allows to interpret more profoundly some phenomenons, notably: the equivalence of a Pauli principle in stationary TE-TM states, the notion of non-locality in a radiation field, a modified form of the De Broglie analysis and the notion of wave-paquets in finite-spacetime.
\end{abstract}

\vskip 10 pt

\noindent
{\bf 1. Introduction.} 

We've shown in a previous paper \cite{kn:AT} that the behavior of an electromagnetic field, developed in TE-TM modes, inside a metallic thermostatic cavity, allows us to analyze how an equilibrium is reached within this system by keeping the finite-spacetime interaction, introducing mode coupling according to Marcuse analysis \cite{kn:Mar} (Power coupled Theory), and analyzing it through short-circuited guided propagation method. We then obtained very interesting results, notably:

\begin{enumerate}
	
	\item The Planck constant ($\hbar$) derived only from classical electromagnetism, in that case, is nothing but the normalized thermodynamic and electromagnetic energy exchange between the electron and the electromagnetic field. This is verified by measures of the complex dielectric constant of good metallic conductors (gold as a reference)
	
	\item The equilibrium in the field of these systems is reached after a chaotic transition and that without knowledge of the initial conditions, only the equilibrium state can be determined.
	
	\item The Planck relation and the Heisenberg inequalities can thus be obtained through classical electromagnetism.
	
\end{enumerate}

Our results showed that classical electromagnetism developed in normal TE-TM modes in a finite-spacetime allows us to represent the blackbody radiation phenomenon without the need for a `photon particle' model, the normalization of thermodynamic and electromagnetic (e.m. field-electron) leading to the determination of Planck constant as, in that case, a physical constant of classical optics and electrodynamics.  Our findings support the point of view according to which the new theory of radiation [] doesn't necessarily, in all cases, need the particle notion of photon but rather an appropriate choice of electromagnetic normal modes to explain the behavior of a radiative system.  We note that the photoelectric effect has already been described without the use of photon (radiation quanta) \cite{kn:Beck}\cite{kn:Lamb}.
 
Following \cite{kn:AT}, the multimodal coupling, an overlooked phenomena, appears as the fundamental notion which explains how the equilibrium is attained in the blackbody radiation and $\hbar$ being, at equilibrium, a normalization constant of the complete set of orthonormal functions used in this classical electrodynamics finite-spacetime approach. The spacetime analysis in quantum physics, which does not use, to begin with, the e.m. limit conditions of the field between the two topological domains of the isolated system (interior and exterior), had developed a genius ``a priori" normalization (Planck-Einstein relation). The QED theory, which can explain with excellent precision experiments, for example, Lamb displacement, hydrogen hyperfine structure, and serves to analyze with success many other interactions, make use of particle-like (photon) in a spacetime paradigm, and had to resolve a renormalization problem using a complex mathematical process. Fig. 1 and fig. 4 of \cite{kn:AT} represent the comparative elements of both approaches, quantum physics and classical electrodynamics we use here.
It is not our goal, in this paper, to interpret all the possible links between classical electrodynamics and quantum physics, but to show that Maxwell equations, analyzed in finite-spacetime, can be used, in part, as a pedagogical paradigm of quantum physics. It will lead to develop and remodel it in a way to escape the \textit{a posteriori} of the spacetime limit conditions now encountered in quantum physics. Electrodynamics is then viewed more as a part of the foundation of quantum physics, rather than an asymptotic formulation of the latter. That paper will not derive new development in that field of physics, but present a new pedagogical approach using, in the cases analyzed, the completeness of the Maxwell theory, after having shown in \cite{kn:AT} the normalization function of $\hbar$ related to a complete set of orthonormal function in the blackbody problem, which also leads to a classical electrodynamic derivation of uncertainty relations.

In the current paper, using this new pedagogical approach, we analyse other characteristics of TE-TM modes and states that are most relevant with respect to the new theory of radiation, namely: the quantification of angular motion, an equivalent of a Pauli principle for these states, a modified formulation of the De Broglie analysis characterizing the notion of wave paquets, the related notion of decoherence, the link between this theory and Shcr\"{o}dinger formalism and finally two cases of application, which will be described in Addendum B.  
The analysis presented here will show that a finite-spacetime quantification, leading to a development of the electromagnetic radiation field (information) in TE-TM states can, in part, serve as another pedagogical approach to quantum physics.

\vskip 10 pt

\noindent
{\bf 2. Angular momentum and equivalent Pauli exclusion principle.}

Having, in the past paper \cite{kn:AT}, derived the Planck constant and the Heisenberg inequalities through classical electrodynamics in normal TE-TM modes in finite-spacetime, the authors will show that the angular momentum quantization follows directly from the energy quantization and momentum analysis of circular TE-TM modes in classical physics. The Pauli Exclusion Principle (rule) will be then described as emerging from classical physics and a more appropriate ``image" of the ``spin" notion, in information processing, will be presented. Those two important results in classical physics about quantization will bring the arousing question: ``Is quantization emerging, in part, from classical electrodynamics analysis in finite-spacetime?".

\vskip 10 pt

\noindent
{\bf 2.1. Angular moment quantization.}
In any metallic empty guide, the electric and magnetic energies are equal. Furthermore,  when the electric field amplitude is zero locally, the magnetic field amplitude reaches a maximum since:

$$
c^2\left(\nabla \times \vec{B}\right)=\frac{d\vec{E}}{dt}.
$$

In the case of the propagation of circular TE-TM modes, we refer the readers to an analysis by Burke and Kapany \cite{kn:Burke} who showed, following Jackson \cite{kn:Jackson}, that in a hollow metallic circular waveguide, the TE-TM circular modes (example $TM_{mp} \pm TM_{-m,p}$) carry a total energy per unit of length (in the direction of propagation m$z$) and a $z$ component of angular momentum per unit of length whose are in the ratio of $-\frac{m}{\omega}$. That result, since we have shown in classical electrodynamics that the electron and field exchange energy per unit of cell is $\hbar \omega$, implied

\begin{equation}\label{eq:eq1}
\frac{\mbox{angular momentum/unit of length}}{\mbox{energy/unit of length}}=\frac{-m}{\omega}
\end{equation}

$$
\frac{\mbox{angular momentum}}{\mbox{unit of length}}=-m\hbar
$$
Thus, by purely classical e.m. approach, and with or without an a priori multiplication of the numerator and denominator by $\hbar$ in (\ref{eq:eq1}), as did Burke and Kapany,  we can assert that fields given by TE-TM modes may be interpreted as single states of quantized e.m. field in a finite-spacetime.

\vskip 10 pt

\noindent
{\bf 2.2. A Pauli exclusion principle for TE-TM e.m. states.}

In an empty metallic cavity, one may think that the magnetic field will give rise to angular momentum since it circulates in closed loop, but this is not accurate. In a stationary TE-TM field, in a cell, the electric field nodes are zero at both ends while the magnetic field node is at the center, i.e. shifted by $\lambda_{g,n}/4$.

\begin{figure}[h]
	\centering
	\includegraphics[width=15cm]{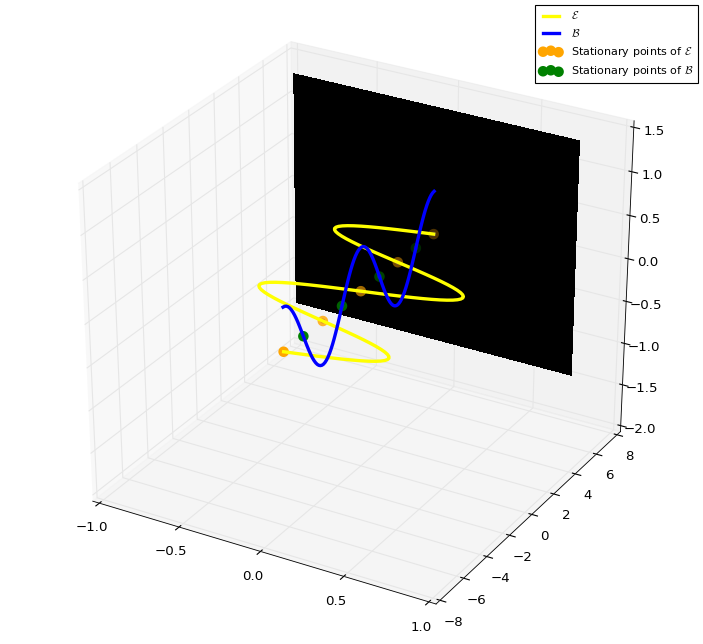}
	\caption{Schematic of transversal components $\mathcal{E}_x$ and $\mathcal{B}_y$ of fundamental field state.}
\end{figure}

Even if we have a null electric field and a maximum magnetic field at the boundaries, the associated angular momentum component $(L_{g,n})$ per cell unit length $(\lambda_{g,n/2})$ cannot be written as $L_{g,n}=\hbar\omega_n=E_m=E_e$, where the last two equalities are for the magnetic and electric energies respectively. Thus $L_{g,n}\not= \hbar\omega$ as the group velocity $v_{gr}\rightarrow c$. The magnetic field node at the center of the cell causes up $+\hbar/2$ and down $-\hbar/2$ angular moments to cancel each other out and for the whole cell, $L_{g,n}=0$. This requires an atom in its ground state, which emits no e.m. radiation, to obey the Pauli exclusion principle as included in solutions of the relativistic expression of Schr\"{o}dinger's equation (e.g. Dirac equation). That image is more pedagogical than the 'spin image' of classical quantum physics.
A perturbation that modifies the stationary state of a system thus involves, locally, a minimum of two cells (electric and magnetic) of dimensions $\lambda_{g,n/2}$ being separated by $\lambda_{g,n/4}$. In terms of the amplitude functions and energy, the field components thus appear locally as a $\hbar$ normalization for the $\vec{\mathcal{E}}$ field and the two components of $+\frac{\hbar}{2}$ and $-\frac{\hbar}{2}$ (up and down) for the $\vec{\mathcal{B}}$ field. So the ``spin effect", in a finite-spacetime, will be subject to those two radiation components. This is the general case when interaction zones (source and observation) obey the Sommerfeld radiation condition $r,r'\gg\lambda$. We can thus establish the foundations of quantum physics on the finite-spacetime aspect of isolated systems and on the TE-TM states associated to the e.m. field (information) developed in these modes individually satisfying the limit conditions.
The application of limit conditions necessarily leads to solving equations leading to eigenvalues and thus to the quantification of energy and its associated motion quantities. The notion of quantification is linked to any finite-spacetime and characterize any information (radiation) which we can obtain from it, according to field theory.
The e.m. stationary model thus implies that the exchanges of energy and angular motion quantities will be linked to discreet values relatively translated by $\lambda_{g,n/4}$.
Finally, note that the notion of finite-spacetime for intervals related to  $r,r'<\lambda_g/2$ refer to evanescent modes. These modes, as indicated \cite{kn:AT}, while orthonormal, have a particularity that is very important with respect to quantum physics: a progressively evanescent mode of index $i$ can exchange energy with the corresponding regressively evanescent mode. The exchanges of energy quantities and motion quantities in a small dimensional interval $(<\lambda_{g/2})$ between a e.m. field developed in TE-TM and particles thus establish a particle-antiparticle type relationship and their analysis should follow the quantum electromagnetism model (QED). Analysis by e.m. TE-TM sates can thus formalize the mathematical and asymptotic  aspect of extreme intervals and avoid, among other things, the renormalization process met in QED, process of which the self consistence has not been shown mathematically \cite{kn:Feyn}. The finite-spacetime model exclude de facto the zero and infinite intervals between events and this, in the relativistic sense. The first exclusion being related to the finite aspect of the universe, and the second to the link established between a source and a probe by the e.m. tunneling effect. The development in TE-TM states and modes implying a phenomenon of inter-modal dispersion, we must now establish a relationship between these and the notion of wave paquets in quantum physics.

\vskip 10pt

\noindent{\bf 3. TE-TM modes and states and wave paquets.}

What is the correspondence between the analysis by TE-TM states and modes and the one using the monochromatic wave paquets?

\vskip 10pt
\noindent{\bf 3.1. Qualitative analysis.}

We can, as a first step, present an approximate solution to these problems. The TE-TM modes (or states) are represented individually by two plane waves propagating at $\pm \theta_n$ of a privileged axis and satisfying the limit conditions of the finite-spacetime which is being considered. As for the monochromatic wave paquet under the Gaussian envelop of quantum mechanics, we can represent them using three principal components. The central component of wave paquet acting as a carrier and containing no information (by itself) we can, relatively to the description by TE-TM modes, choose the lateral bands at $\pm \frac{\delta k}{2}=2\sqrt{a}$ of impulsion

\begin{equation}\label{eq:eq2}
\Phi(k)=\frac{A}{\sqrt{4\pi a}}e^{-(k-k_0)^2/4a}
\end{equation}

as elements of relevant comparison (information) with the constructive waves of a TE-TM mode. We obtain, because

\begin{equation}\label{eq:eq3}
k_g^2=k_o^2-k^2_{c,eq}
\end{equation}

and that generally, $k_c\ll k_o$ and $k_g\rightarrow k_0$, (3) becomes
\begin{equation}\label{eq:eq4}
k^2_{c,eq}=k_o^2-k_g^2=(k_o+k_g)(k_o-k_g)= 2k_0\Delta k
\end{equation}

\begin{equation}\label{eq:eq5}
k_{c,eq}\approx \sqrt{2k_0\Delta k.}
\end{equation}

If the wave paquet has a quality factor $Q$ given by
 
\begin{equation}\label{eq:eq6}
Q=\frac{\omega_0}{\Delta \omega},
\end{equation}

we can write (5) as

\begin{equation}\label{eq:eq7}
\lambda_{c,eq}=\frac{\lambda_0}{2}\sqrt{2Q} \mbox{ and } b_{c,eq}=\lambda_0\sqrt{Q}
\end{equation}

where $b_{c,q}$ represents the transversal dimension equivalent for the mode distribution.  
An electron perturbed, for example, by an e.m. wave of $\omega$ and $\Delta\omega$ cannot differentiate between the two following situations: it is perturbed by a source  of $\omega$ and $\Delta \omega$ or by a mode of a finite e.m. field of transversal dimension $b=\lambda_0 Q^{1/2}$.  The notion of isolated system thus implies an analysis requiring a finite spacetime and must respect the vectorial aspect of $\vec{k}$, the wave vector and the transversal resonance conditions \eqref{eq:eq3}.

\vskip 10pt

\noindent{\bf 3.2. Analysis by analogy of modified De Broglie relation.}

A more precise analysis of the relationship between TE-TM modes and states and the wave paquets function must start with the following quantum mechanic relationship:

\begin{equation}\label{eq:eq8}
E^2=p^2c^2+m_0^2c^4
\end{equation}

\footnotemark which relates the total energy and the motion quantity of a particle (even relativistic).

\footnotetext[1]{This relationship implies that, measured in a given reference frame, the total relativistic energy of an isolated system remains constant. Observe that the  isolated system implies a topologically finite spacetime. The fields are contained in this isolated space (finite-spacetime) by the so called hard reflexions or by refraction on caustics (respective phase jumps of $\pi$ and $\pi/2$).} 

Feynman \cite{kn:Feyn2} has already noted a pedagogical similarity between relation (\ref{eq:eq7}) and these of TE-TM waves in rectangular hollow metallic guides, i.e. the following relationship:

\begin{equation}\label{eq:eq9}
k_g=\sqrt{\frac{\omega^2}{c^2}-\frac{4\pi^2}{\lambda_c^2}}.
\end{equation}

Indeed substituting in (\ref{eq:eq9}) as in quantum mechanic $E=\hbar\omega$ and $\hbar/\lambda=\hbar k$, he obtains

\begin{equation}\label{eq:eq10}
k=\sqrt{\frac{\omega^2}{c^2}-\frac{m_0^2c^2}{\hbar^2}}.
\end{equation}

He qualifies this result as interesting! Thus we modify, in this paper, the de Broglie analogy by introducing an analog  being

\begin{equation}\label{eq:eq11}
\lambda_c=\frac{\hbar}{m_0c}.
\end{equation}

corresponding to a cutoff wave length (Compton type; inertia mass related in quantum mechanics) for the particle. We thus obtain the following relationship:

\begin{equation}\label{eq:e12}
v_{ph}=\frac{\omega}{k}=\frac{\omega}{\sqrt{\frac{\omega^2}{c^2}-\frac{m_0^2c}{\hbar^2}}}=c\left[1-\frac{m_0^2c^2}{\hbar^2}\right]^{-1/2}.
\end{equation}

The group velocity according to the TE-TM analysis takes the form

\begin{equation}\label{eq:eq13}
v_g=\frac{d\omega}{dk}=c\left[1-\frac{\lambda_0^2}{\lambda_c^2}\right]^{1/2}=v.
\end{equation}

Because

\begin{equation}\label{eq:eq14}
m=m_0\left(1-\frac{v^2}{c^2}\right)^{-1/2},
\end{equation}

we obtain the relationship

\begin{equation}\label{eq:eq15}
k_g^2=k_0^2-k_c^2
\end{equation}

where

$$
k_c=\frac{m_0c}{h}=\frac{2\pi}{\lambda_c}\mbox{,  } k_g=\frac{mv}{h}=\frac{2\pi}{\lambda_g}\mbox{, } k_0=\frac{mc}{h}=\frac{2\pi}{\lambda_0}.
$$

The de Broglie analogy associated with a Compton cutoff wavelength\footnotemark \space is thus necessary in order to satisfy relativity and the vectorial nature of $\vec{k}_g$. The relation (\ref{eq:eq15}) indicates that a development in TE-TM states would then be appropriate in the case of finite-spacetime.  
In summary, a development in TE-TM modes or states satisfying the transverse resonance relationship associated to the finite-spacetime considered  will rely on the analogical wave with group speed $v_g=v$ that will be coupled to the particle during its perturbation (event). This analysis here gives a physical meaning to the quantum mechanic relationship \cite{kn:Rauch},

\begin{equation}\label{eq:eq16}
\frac{\omega}{k}=\frac{c^2}{v}=v_{ph}=c\sqrt{1+\frac{m_0^2c^2}{\hbar^2k^2}} ,
\end{equation}

indicating that phase speed is always superior to $c$. This observation links this last relation to the finite aspect of the isolated system. Thus it precises the notion of a wave function as ``an expression and a summary of the available information for an observer" \cite{kn:AT}.

\footnotetext[2]{Not taking this particularity into account in the De Broglie relation can lead to interpreting group velocity as being superior to $c$.}

\vskip 10pt
\noindent{\bf 4. Multimodal coupling in classical optics and electrodynamics and decoherence in quantum physics. }

At the macroscopic level, the notion of multimodal coupling related to the modeling of an isolated system (finite-spacetime), which, per se, leads to the quantization of the information received by an observer, explains well the notion of decoherence. The phenomena of equilibrium attainable in blackbody radiation described and explained within the classical electrodynamics theory by the authors corresponds to a ``typical" example of decoherence.

\vskip 10pt

The multimodal coupling (power coupling) plays a dominant role, being generated and synchronized by different physical phenomena (matter oscillators).

\vskip 10pt

The approach by finite-spacetime implied that the ``average power" of radiation involved must be privileged over the ``average energy" i.e. the use of ``Power coupled theory of Marcuse", thus keeping together spacetime in the analysis. Not doing that, as an example, in the ultraviolet catastrophe in the blackbody radiation problem, can lead to incomplete analysis of an electrodynamic problem. Then, when Henri Poincarr\'e, cited in [10], showed ``that the average energy of an oscillator in matter must be the same as the average energy of an electromagnetic oscillator", he refers to ``static equilibrium" of the isolated system, while the normal one has to be ``statistical" and related to the ``average power" and ``tolerated noise" in that system.

\vskip 10pt

The process of decoherence is irreversible in both classical electrodynamics and quantum physics as noted in reference \cite{kn:AT, kn:Omnes}

\vskip 10pt

Also noted by Omnes \cite{kn:Omnes}, ``decoherence is certainly the most efficient and rapid effect to exist in macroscopic physics". So is multimodal power coupling in classical electrodynamics.

\vskip 10pt

Classical electrodynamics and quantum physics both assert that decoherence in dissipative systems always bring down in frequency the received information \cite{kn:Jackson}. However, from the results we obtained in ref. \cite{kn:AT} (see fig. 1), that very statement has to be specified. In fact, in a multimodal system, decoherence can also force a group of lower frequency modes to transfer energy to higher frequencies when reaching thermodynamic equilibrium.

\vskip 10pt

To resume, as noted in \cite{kn:Omnes}, one cannot say that the notion of decoherence in classical physics is emerging from quantum physics. Decoherence is a result, in finite-spacetime, of e.m. modes and states coupling and smearing of information an observer can obtain from an isolated system.

\vskip 10pt
\noindent{\bf 5. TE-TM modes and states and the Schr\"{o}dinger's equations. }

Which analysis of current quantum physic corresponds the most to the one described in our paper?  The description of a remarkable analogy between Schr\"{o}dinger's equation and guided wave propagation in optical fiber with a distribution of refractive index $n(r)$ answers this question. Okoshi \cite{kn:oko} demonstrates that it is then possible to obtain two independent solutions from the electromagnetic wave equations with radial and azimuthal field components of the form:

\begin{equation}\label{eq:eq17}
F(r)\left\{\genfrac{}{}{0pt}{}{\cos(v\phi)}{\sin(v\phi)}\right\}\exp(j\beta z)
\end{equation}

Where $F(r)$ is solution to the equation:

\begin{equation}\label{eq:eq18}
\frac{d^2F}{dr^2}+\frac{1}{r}\frac{dF}{dr}+\left(n^2(r)k_0^2-\beta^2-\frac{v\pm 1}{r^2}\right)F=0.
\end{equation}

Combining Okoshi EH and HE solutions, the uniform polarization LP modes are obtained for which the non zero transverse Cartesian component of the field has the same form as eq. (\ref{eq:eq18}) with $F(r)$ solution to the equation.

\begin{equation}\label{eq:eq19}
\frac{d^2F}{dr^2}+\frac{1}{r}\frac{dF}{dr}+\left(n^2(r)k_0^2-\beta^2-\frac{v^2}{r^2}\right)F=0.
\end{equation}

Okoshi posits $\hat{F}=\sqrt{r}F$ and transforms eq. (\ref{eq:eq19}) into

$$
\frac{d^2\hat{F}}{dr^2}+\left[E-V(r)\right]\hat{F}=0
$$

where $E=n_1^2k_0^2-\beta^2$and $V(n)=n_1^2+n^2(r)k_0^2+(v^2-1/4)/r^2$, $n_1$ being the refractive index maximum in the optical fiber. One can recognize Schr\"{o}dinger's equation with $E$ being the energy and $V(r)$ the potential.  The solutions are thus three fold:

\textit{a)} If $0<E<V_{\infty}$ then $n_2k_0<\beta<n_1k_0$ i.e. the guided wave mode cases. 

\textit{b)} and \textit{c)} If $E<V_{\infty}$ then $\beta<n_2k_0$ i.e. leaky waves are obtained by tunnel effect \textit{(b)} or refraction \textit{(c)}.

Remark that $\hat{F}(r)$ in the asymptotic case  $(\beta_g\rightarrow \beta_0)$ will tend toward functions $\frac{\sin(kr)}{r}\left(\frac{\cos(kr)}{r}\right)$ and that fields will appear  of the TE-TM type with slow modulation.

Let us mention that this analogy between the Schr\"{o}dinger equation and TE-TM modes has also been applied to the study of resonance in a dielectric sphere in Addendum A.  The most remarkable analogies are therefore related to phenomena of guided propagation that appeal to the solution of eigenvector and eigenvalue equations. One must preserve a finite-spacetime link to analyse them.

This brief description indicates that the quantification of e.m. field in finite space-time is better adapted to that of Schr\"{o}dinger's picture, while the formalism of e.m. field quantification in free space time (free e.m. field photon approach) respects Heisenberg's formalism. Finally, the reader will find in addendum B a presentation of two cases in need of modal approach; the constant of fine-structure and the notion of single photon.

\vskip 10pt
\noindent{\bf 6. Conclusion and prospectives.}

Gauge invariance and transversality of the e.m. field are usually considered in spacetime. In doing so, it leads to the plane wave representation of the wave aspect of the particle-like photon. We did not use, in this paper, Coulomb and Lorentz gauges. The reader will find an analysis of problems raised by a spacetime quantization in the references \cite{kn:Weinberg, kn:Iz}.

We have instead proposed, as another pedagogical approach, a quantification of the e.m. field better adapted to isolated systems i.e. a quantification of e.m. field in finite-spacetime. The development of e.m. field in TE-TM states and modes in an isolated system leads to a quantization which does not explicitly rely on the notion of photon nor the general application of gauge.  This quantization obeys the limit conditions of a finite-spacetime which follow from the solution of an eigenvalue equation and allows to interpret more profundly some phenomena, notably these discussed in \cite{kn:AT} and in this paper: the establishment of an equilibrium in a black body radiation; the equivalence of a Pauli principle in stationary TE-TM states; a modified form of the De Broglie analysis; the uncertainty relations and wave-paquets in finite-spacetime. The analysis by TE-TM states avoids problems associated with choosing a gauge because the elements of the complete and orthonormal states must satisfy the limit conditions of a finite-spacetime and, doing so, leads to the resolution of an eigenvalue problem and to the quantization of fields. In summary, after a period of analysis in space and time, followed by the analysis in spacetime, we introduced, in this paper, the analysis in finite-spacetime which shows that the quantization is linked to information that we can obtain from an isolated system. We then relate more closely the quantification on electromagnetism and give to a part of quantum mechanic a pedagogical approach based on the Maxwell theory.

\vskip 10pt

One cannot say as Wheeler did ``Why Quantum Mechanics?" or try to justify the emergence of classical physics from quantum physics, but can accept the relation $classical$  $physics \rightleftharpoons quantum$  $physics$ which respects the Hilbert statement, cited in \cite{kn:Omnes}; ``The same theory cannot contain two categories of axioms, some of which belong to classical physics and others to quantum physics". This logical statement can serve for further analysis. In the first paper \cite{kn:AT}, the authors, using a finite spacetime approach, reconciliated classical physics and quantum physics in regards of the notion of quantization of energy, uncertainty relations, and, in part, of ``decoherence" in multimodal coupling.

\vskip 10pt

\noindent{\bf 6.1. Pauli principle (rule)}

In the present paper, one realizes that in classical (macroscopic-e.m) a rule (a quantum axiom), the Pauli exclusion principle applies well if one uses, as informative results, the development of e.m field (radiation) in a complete orthonormal base, the TE-TM e.m modes and states in finite-spacetime (section 2). It leads to a more pedagogical approach ``image" that the usual ``spin" image of the classical quantum physics only in space-time and open the domain to further analysis since that principle stands in finite-spacetime. Note, as exemple, that the notion of locality refers to two different cells for the magnetic states within an electric cell. Further studies are needed, in particular these related to the phenomena of macroscopic mass interference and diffraction.

\vskip 10pt

\noindent{\bf 6.2. The notion of oscillators}

In the history of quantum mechanics, three kinds of oscillators are cited as having a role to play, citing Omn\`es \cite{kn:Omnes}; ``The first type was abstract : a purely theoretical construct that was used by Max Planck as a tractable model of matter. The second included electromagnetic oscillators, also theoretical notions, which provided a convenient representation of radiation. Their properties were essentially equivalent to Maxwell's equations from the standpoint of dynamics and, when  quantized, they became ``photons''. The third species of oscillators was the only physically manifest one because it consisted of elastic vibrations of crystal. They became ``phonons" in their quantum version ". In the approach used in our first paper \cite{kn:AT} and this one, the TE-TM e.m modes and states development of radiation fields and quantization in terms of the notion of finite-spacetime, leads to the unification of the first two. It also presents an operational relation between the physical and theoretical oscillators via the power multimodal coupling theory since it needs the variable for time, $t$, to be involved in finite spacetime.

\vskip 10pt

\noindent{\bf 6.3. Completeness}

The approach proposed, as far as the completeness of the theory is concerned, rest upon a ``complete and orthonormal development and, contrary to the photon-like particle one, does not need a process of normalization and the ``a posteriori" aspects of the limit conditions which are encountered in Quantum Physics. As it is shown in section 3.2 of this paper, all modes of the development are linked to $v$ and one does not need to assume a process of collapsing of the wave paquet, nor the notion of pilot waves and hidden variables. 

\vskip 10pt
\noindent{\bf 7. Acknowledgments.}

This work was supported by the National Sciences and Engineering Research Council of Canadanand the Government of Quebec. We also acknowledge the precious help of Professor Claudine Ni Allen in regard of her contribution to section 2 and the helpful contribution of Jean-Fran\c cois Tremblay to the editing of this paper.

\vskip 20pt

\centerline{\textbf{\textit{Addendum A}}}
\centerline{\bf Geometrical justification of the TE-TM development}

\vskip 10pt

Let us mention that this method of analysis, an analogy between the Schr\"{o}inger equation and the TE-TM electrodynamics, has also been applied to the study of resonance in a dielectric sphere under the name of morphology-dependent resonance (MDR's) . The reader can refer to a review article written by Hill and Benner \cite{kn:Hill} and to an article by B.R. Johnson \cite{kn:Johnson} who developed this theory in analogy with the theory of quantum-mechanical shaped resonance. This type of theory is based on the description of the fields inside and around a dielectric sphere, which obey Maxwell's equations

\begin{equation}\tag{A}
\nabla \times (\nabla \times \vec{E}) + \frac{\varepsilon(r)}{c^2} \frac{\partial^2 \vec{E}}{\partial t^2} = 0
\end{equation}

where $c$ is the speed of light in empty space. One represents the field $\vec{E}$ in a spectral development from, i.e,

\begin{equation}\tag{B}
\vec{E} = \int_0^\infty \vec{e} (r) e^{c \omega t} d\omega
\end{equation}

and (20) becomes

\begin{equation}\tag{C}
\nabla \times \nabla \times \vec{e} - k^2 \varepsilon (r) \vec{e} = 0
\end{equation}

where $k = \omega / c$, the wave vector.\newline
This equation can be developed in TE-TM, ``since $\nabla \cdot (\varepsilon \vec{e}) = 0$", leading to

\begin{equation}\tag{D}
\vec{e} = \sum_{\nu, m}^{} \frac{1}{r} \Big[\Psi_{TE} Y_{\nu,m} + \frac{1}{\varepsilon (r)} \nabla \times (\Phi_{TM} \vec{Y}_{\nu,m}) \Big]
\end{equation}

where $Y_{\nu,m}$ is a spherical vectorial function with $\nu$ as its angular number and $m$ as its magnetic number. The radial distribution of the field for TE modes, i.e for a dielectric cavity, either spherical or cylindrical, has the following form:

\begin{equation}\tag{E}
\frac{\partial^2 \psi}{\partial r^2} + \Big [k^2 \varepsilon (r) - \frac{\nu(\nu+1)}{r^2} \Big ] \psi = 0
\end{equation}

This way we get back to Okoshi's approach described in section 5. As for the well-known coupled-mode theory \cite{kn:Mar}, its consistency is well established when derived from two Maxwell wave equations without using approximations in finite-spacetime \cite{kn:MFloRTre1}\cite{kn:MFloRTre2}. The direct derivation of the coupled-mode theory thus proposed, clarifies the role of the slowly-varying amplitude approximation. It is also reasonable to keep the electric and magnetic fields in the description, not only for energy reasons but also for a ``locality'' consistency, in virtue of section 2 of this paper, since the propagation time averaged power is given by the integrated Poynting vector, over an infinite cross-section and is determined by the product of electric and magnetic fields.

\vskip 20pt

\centerline{\textbf{\textit{Addendum B}}}
\centerline{\bf Two other cases in need of a finite-spacetime paradigm}

\vskip 10pt

The two other cases of interest, briefly presented below, refer to the constant of fine-structure and the notion of single photon. The preliminary analysis shows they are in need of the paradigms proposed in this paper.

\vskip 10pt

\centerline{\bf \S 1 The fine-structure constant}

\vskip 10pt

Let's shortly consider, for the finite-spacetime paradigm, what is called in classical quantum physics the fine-structure constant, which is:

\begin{equation}\nonumber
	\gamma = \frac{e^2}{4 \pi \varepsilon_o c \hbar} = \frac{1}{137,...} .
\end{equation}

This finite-spacetime relation, which uses the transformation relation $\mu_0 \varepsilon_0 c^2 = 1$ and the notion of characteristic impedance related to the transversal components of TE-TM fields, which are \cite{kn:Jackson}

\begin{equation}\nonumber
	(Z_{car})_n = \begin{cases}
	\text{ \Large $
	\frac{k_{g,n}}{\varepsilon \omega} = \frac{k_{g,n}}{k_0} \sqrt{\frac{\mu}{\varepsilon}}
	$}
	, & \text{(TM)}\\
	
	\text{ \Large $
	\frac{\mu \omega}{k_{g,n}} = \frac{k_0}{k_{g,n}} \sqrt{\frac{\mu}{\varepsilon}}
	$}
	, & \text{(TE)}
	
	\end{cases}
\end{equation}

can be written, in the case of propagation in free-space and the case where $k_{g,n} \rightarrow k_0$, as

\begin{equation}\nonumber
 \gamma \simeq \frac{Z_0 e^2}{4 \pi \hbar}
\end{equation}

This approximation implies that said fine-structure constant, in spacetime, only links charges to the modal characteristic impedance of the transversal e.m. field and $\hbar$, the normalization factor of the energy exchange between electric charges and the e.m. field \cite{kn:Jackson, kn:Bishop}. However, as part of the finite-spacetime paradigm, one needs to be more rigorous and to generalize this relation using, as a starting point, characteristic impedances $(Z_{car})$. It is then normal that, in quantum physics, one came to characterize $\gamma$ as dependent of the ``energy scale of the studied process as well as of the normalization procedure detail" \cite{kn:Bellac}, considering that in classical electrodynamics the modal notion is linked to frequency and energy. The authors will not discuss this constant further but they note that, in finite-spacetime, the ``vacuum fluctuation" notion is unappropriated if we must respect the $c^2 \mu_0 \varepsilon_0 = 1$ transformation. The observed fluctuation should then be linked to the mode type and the concerned coupling, not attributed to a ``vacuum state", the latter being, essentially, going back to the ``ether problem''.

\vskip 10pt

\centerline{\bf \S 2 The single-photon notion}

\vskip 10pt

The Abraham-Minkowski dilemma will be used to illustrate that the single-photon notion is unappropriated in finite-spacetime and that a TE-TM development of the e.m. fields is, in this case, necessary.

\vskip 10pt

The reader will find in references \cite{kn:Boyd} two analysis of this dilemma where the authors conclude that both representations of the momentum density of light going through a material with $M$ as its mass, $\vec{E} \times \vec{A}$ according to Abraham and $\vec{D} \times \vec{B}$ according to Minkowski, are correct. However, for a planar wave (photon), they lead to two different results:

\begin{equation}\tag{A}
	\vec{P}_{Abr.} = \frac{\vec{p}_0}{n_m} \text{\qquad and \qquad} \vec{P}_{Min.} = \vec{p}_{\gamma} n_m
\end{equation}

where $\vec{p}_0 = \hbar \vec{k}_0$ is the momentum in free space and $n_m$ is the refraction index in the considered material. The relations in (A) correspond to a spacetime analysis. Do note, however, that S. Barret \cite{kn:Barret}, who came to the same conclusions, report that in a dispersive environment where the phase index $n_p = \frac{ck}{\omega}$ and the group index $n_{g_s} = c(\frac{d\omega}{dk})^{-1}$ are different, experiments respectively measure $P_{min} = \frac{\hbar \omega n_p}{c}$ and $P_{Abr} = \frac{\hbar \omega}{cn_{g_2}}$. He adds, furthermore, that one should do a modal analysis.

\vskip 10pt

The cited authors are, usually, only interested in the behavior of a single photon (a plane monochromatic wave) and thus neglect the finite aspect of the considered spacetime \footnotemark . The characterization of total momentum is not that simple since it should use a TE-TM development of the e.m. field. The total momentum would related to the momentum associated with the different modes, in a first approximation, if we take into account that a mode is equivalent to a "wave paquet" as described in section 3.

\footnotetext[3]{Note that, in general, we refer to ``single photon wave paquets" and ``single photon states" (A, Aspect and P. Granger "Wave-particle duality for single photons" Hyperfine Interactions 37 (1987) p 3-18)}

\end{document}